\documentstyle[eqsecnum,aps,epsf]{revtex} 
\newcommand{\postscript}[2] {\setlength{\epsfxsize}{#2\hsize}
\centerline{\epsfbox{#1}}}

\begin{document}
\twocolumn[\hsize\textwidth\columnwidth\hsize\csname@twocolumnfalse\endcsname

\title{Bose-Einstein condensation thermodynamics  of a trapped  gas with
attractive interaction}

\author{Sadhan K. Adhikari}
\address{Instituto de F\'{\i}sica Te\'orica, Universidade Estadual Paulista,
01.405-900 S\~ao Paulo, S\~ao Paulo, Brazil\\}

\date{\today}
\maketitle
\begin{abstract}

We study the Bose-Einstein condensation of an interacting gas with
attractive interaction confined in a harmonic trap using a semiclassical
two-fluid mean-field model. The condensed state is described by
converged numerical solution of the Gross-Pitaevskii equation.  By solving
the system of coupled equations of this model iteratively
we obtain
converged results for the temperature dependencies of the condensate
fraction,  chemical potential, and internal energy  for
 the Bose-Einstein condensate of $^7$Li
atoms.


{\bf{{To appear in Physica A}}
} 

\end{abstract} 

\vskip1.5pc] 

\section{Introduction}

There has been recent experimental observation of Bose-Einstein (BE)
condensation in dilute interacting bosonic atoms of $^{87}$Rb
\cite{1a,1b}, $^{23}$Na \cite{2}, $^7$Li \cite{3,4}, and $^1$H \cite{5}
employing magnetic traps at ultra-low temperatures. The interaction among
the atoms could be either attractive or repulsive. Although, in both cases
the condensate is well-described by the Gross-Pitaevskii (GP) equation
\cite{6a,6b}, the nature of BE condensates in these two cases is of
entirely different nature. In the repulsive case the number of atoms in
the condensate can grow without bound, whereas this number is limited by a
upper bound in the case of attractive interaction \cite{7a,7b,7c}. Of the
experimentally observed cases of BE condensation one has repulsion for
$^1$H, $^{23}$Na, and $^{87}$Rb atoms and attraction for $^7$Li atoms. The
existence of a maximum number of condensed atoms in the case of $^7$Li has
been noted experimentally and is consistent with the prediction of
theoretical analysis based on the GP equation \cite{4,7a,7b,7c,8a,8b}.  In
the attractive case the GP equation has no solution for the number of
atoms larger than a critical number.

The GP equation is a nonlinear Schr\"odinger equation and it is tedious to
find its converged numerical solution \cite{8a,8b,9a,9b,10a,10b}.  For the
repulsive
case, an approximate solution scheme of this equation, such as the one
based on the Thomas-Fermi approximation, has frequently been used for a
qualitative description of the condensate \cite{11}. However, no such
approximation scheme is known for the attractive case which
requires an exact numerical solution of the GP equation. This makes the
theoretical study of this case a more challenging  task.

There have been several comprehensive studies on the temperature
dependencies of the thermodynamic observables in the case of repulsive
interaction using mean-field two-fluid models
\cite{11,13,13a1,13a2}. One such
mean-field scheme is provided by the so-called Popov approximation
\cite{14} and has been considered by several authors
\cite{13,13a1,13a2,15,16}.
The
physical ingredients of these mean-field models
\cite{11,13,13a1,13a2,15,16} are
quite similar and they lead to similar numerical results in the case of
weakly repulsive interatomic interactions.

We shall use a two-fluid mean-field model \cite{11} to study the
temperature dependencies of the thermodynamic observables of the
condensate.  For a condensate composed of 40000 trapped $^{87}$Rb atoms
the perturbative solution of the system of equations of this model
converged rapidly and provided a satisfactory account of the condensate
fraction, internal energy, and specific heat in agreement with experiment
\cite{1a,1b}. It was also found that the lowest order solution already
provided a very good approximation. Later the same model has been used in
one and two space dimensions \cite{12a,12b}.

The above mean-field two-fluid model \cite{11} is used in this work for a
theoretical description of the BE condensation thermodynamics in the case
of attractive interaction appropriate for $^7$Li. Depending on the
strength of the attractive potential, the condensate in the attractive
case may consist of a few thousand atoms confined by the trap potential.
For a fixed trap, the maximum number of atoms in the BE condensate with
attractive interaction is inversely proportional to $|a|$ \cite{15}, where
$a$ is the scattering length of two atoms.  As the temperature is lowered
below the critical temperature $T_0$ of BE condensation, the condensate
starts to form and finally at 0 K all the available atoms (limited by the
maximum number mentioned above) form the condensate in the present model.

The condensate wave function in the present model is described by the GP
equation. In the repulsive case usually some approximate solutions of the
GP equation are used \cite{11,13,15}. Although we shall be using the
iterative solution of the system of equations of the mean-field model, we
shall employ a converged numerical solution of the GP equation in the
present attractive case. As the GP equation is a nonlinear one, this
amounts to a nontrivial modification of the calculational scheme.

The plan of the paper is as follows. In Sec. II we present the mean-field
two-fluid model. In Sec. III we discuss the numerical scheme for its
solution and present numerical results for $^7$Li. Finally, in Sec. IV we
present some concluding remarks.

\section{Mean-field Two-fluid Model}

We consider a system of $N$ bosons with attractive interaction at
temperature $T$ under the influence of a trap potential. The condensate is
described by the following GP equation for the wave function $\Psi({\bf
r})$ with eigenvalue $\bar \mu$ \cite{6a,6b,11,15}:
\begin{equation}\label{1} \left[
-\frac{\hbar^2\nabla^2}{2m}+V_{\mbox{ext}}(r)  +2gn_1({ r})+g\Psi^2({\bf
r})-\bar \mu \right]\Psi({\bf r})=0.  \end{equation} Here
$V_{\mbox{ext}}(r)\equiv m\omega^2r^2/2$ is the spherically symmetric
harmonic-oscillator trap potential, $g\equiv 4\pi\hbar^2a/m$ the strength
of the interatomic interaction, $m$ the mass of a single atom, $\omega$
the angular frequency, $a$ the atom-atom scattering length, $\mu \equiv
\bar \mu - \bar \mu _0$ is the chemical potential, where $\bar\mu_0$ is
the eigenvalue of Eq. (\ref{1}) for the harmonic oscillator potential
alone in the absence of interatomic interaction ($g=0$), and $n_1(r)$
represents the distribution function of the noncondensed bosons. Although
in actual experiment \cite{4} the harmonic oscillator trap is not quite
symmetric, the deviation from spherical symmetry is quite small. However,
the converged numerical solution of the GP equation (\ref{1}) for a
nonsymmetric trap is quite complicated numerically and hence for a
qualitative description we consider a spherically symmetric trap in the
present study.  An attractive (repulsive) interaction corresponds to
negative (positive)  values of $a$ and $g$.

The noncondensed particles are treated as non-interacting bosons in an
effective potential $V_{\mbox{eff}}(r )=V_{\mbox{ext}}(r)
+2gn_1(r)+2g\Psi^2({\bf r})$ \cite{17}.  Thermal averages are calculated
with a standard Bose distribution of the noncondensed particles in
chemical equilibrium with the condensate governed by the same chemical
potential $\mu$. In particular the density $n_1(r)$ is given by \cite{17}
\begin{equation}\label{2} n_1(r) = \frac{1}{(2\pi
\hbar)^3}\int\frac{d^3p}{\exp[\{p^2/2m+V_{\mbox{eff}}(r)-\mu\} /k_BT]-1},
\end{equation} where $k_B$ is the Boltzmann constant.  Equations (\ref{1})
$-$ (\ref{2}) above are the principal equations of the present model.

The total number of particles $N$ of the system is given by
\begin{eqnarray}\label{3} N=N_0+\int \frac
{\rho(E)dE}{\exp[(E-\mu)/k_BT]-1}, \end{eqnarray} where $N_0 \equiv \int
\Psi^2({\bf r}) d^3r$ is the total number of particles in the condensate.
The semiclassical density of states $\rho(E)$ of Eq. (\ref{3}) is given by
\cite{17} \begin{equation} \rho(E)=\frac{(2m)^{3/2}}{4\pi^2\hbar^3}
\int_{V_{\mbox{eff}}(r)<E}\sqrt{E-V_{\mbox{eff}}(r)} d^3r. \label{4}
\end{equation} The critical temperature $T_0$ is obtained as the solution
of Eq. (\ref{3}) with $N_0$ and $\mu$ set equal to 0.

The average single-particle energy of the noncondensed particles is given
by\cite{11} \begin{equation}\label{5} \langle E \rangle_{\mbox{nc}}=\int
\frac {E\rho(E)dE}{\exp[(E-\mu)/k_BT]-1}.  \end{equation} The kinetic
energy of the condensate is assumed to be negligible and its interaction
energy per particle is given by $ \langle E \rangle _{\mbox{c}}=(g/2)\int
\Psi^4({\bf r}) d^3r$. The quantity of experimental interest is the
average total energy $ \langle E \rangle = [\langle E \rangle
_{\mbox{nc}}(N-N_0)/2+\langle E \rangle _{\mbox{c}}]/N, $ which we
calculate.

The coupling of the nonlinear equation (\ref{1}) with other equations make
the solution algorithm complicated and it is convenient to express this
system of equation in dimensionless units. This has advantage in the
numerical solution \cite{10a,10b}.  We express energy in units of
$\hbar\omega$, and length in units of the harmonic oscillator length
$a_{\mbox{ho}} \equiv \sqrt{\hbar/(m\omega)}$. We shall consider in this
paper only the spherically symmetric ground state solution of the
condensate with $\Psi({\bf r})=\Psi(r)$. Then the GP equation (\ref{1})
become \begin{eqnarray}\label{6} \left[ -\frac{d^2}{dx^2} +x^2+4\pi \eta
\bar n_1(x)  \pm \frac{2\phi^2(x)}{x^2}-2\bar \alpha \right]\phi({ x})=0,
\end{eqnarray} where $\eta\equiv 4a/a_{\mbox{ho}}$ is the new
dimensionless strength, $x \equiv r/a_{\mbox{ho}}$, $\Psi({ r}) \equiv
\phi(x)/(x\sqrt{\pi |\eta|a^3_{\mbox{ho}}}),$ and $\bar \alpha \equiv \bar
\mu/(\hbar\omega)$. The positive (negative) sign in Eq. (\ref{6})
corresponds to repulsive (attractive) interaction. The dimensionless
density $\bar n_1(x)\equiv a_{\mbox{ho}}^3 n_1(r)$ is defined by
\begin{equation}\label{7} \bar n_1(x) = \frac{1}{2\pi ^2}\int\frac{k^2
dk}{\exp[\{k^2/2+V_{\mbox{eff}}(x)-\alpha\} /t]-1}, \end{equation} where
$k \equiv pa_{\mbox{ho}} /\hbar$, $V_{\mbox{eff}}(x)\equiv x^2/2+2\pi \eta
\bar n_1(x)+2\phi^2(x)/x^2$,   $t \equiv k_B T/(\hbar\omega)$, $\alpha
\equiv \mu/(\hbar\omega)$. For the spherically
symmetric ground state $\alpha =\bar \alpha -\bar \alpha_0 $, where
$\bar \alpha_0 \equiv \bar \mu _0/(\hbar \omega)=1.5$ is the energy
eigenvalue of the harmonic oscillator
potential alone (zero-point energy in units of $\hbar \omega$) in the
absence of interatomic
interaction.  In the present consideration  of the  chemical potential
finite-size effects are excluded \cite{11,13}.
Using  these dimensionless variables the number
equation (\ref{3}) becomes \begin{eqnarray}\label{8}
N=N_0+\int  \frac {\bar \rho(e)de}{\exp[(e-\alpha)/t]-1},
\end{eqnarray}
where $N_0=(4/|\eta|)\int \phi^2(x) dx$, $e \equiv E/(\hbar\omega)$,
and the dimensionless density $\bar
\rho(e) \equiv
\hbar \omega \rho(E)$  is given by
\begin{equation}
\bar \rho(e)=\frac{2\sqrt
2}{\pi}\int_{V_{\mbox{eff}}(x)<e}\sqrt{e-V_{\mbox{eff}}(x)}  x^2   dx.
\label{9}
\end{equation}

The above set of equations (\ref{6}) $-$ (\ref{9}) are solved iteratively
using the converged numerical solution of the GP equation (\ref{6}). The
iteration is started with $\bar n_1(x) =0$ at a definite temperature with
a trial value for the chemical potential $\alpha$. Then Eq. (\ref{6})  is
solved and with its solution the functions $V_{\mbox{eff}}(x)$ and $\bar
n_1(x)$ are calculated. Using these new $V_{\mbox{eff}}(x)$ and $\bar
n_1(x)$  Eq. (\ref{6}) is solved
again and  $\bar n_1(x)$ and $\phi(x)$ are
recalculated. This iterative scheme is continued until final convergence
is
achieved. In each order of iteration we calculate the condensate fraction
$N_0/N$ and  energy $\langle E \rangle$, in addition to the
chemical
potential $ \alpha$.  This scheme is
repeated until convergence is achieved. It is then verified if the number
equation (\ref{8}) is satisfied with this solution. If not, a new trial
value for the chemical potential is employed.
Once the number equation is satisfied  the desired solution is obtained.

Next we present the solution procedure of the GP equation (\ref{6}). The
solution of this equation satisfies the following boundary conditions
\cite{9a,9b} \begin{eqnarray} \phi(0) &=& 0, \hskip 1cm \phi '(0)=
\mbox{constant}\label{10}\\ \lim _{x\to \infty}{\phi(x)}& = & N_C \exp
\left[ -\frac{x^2}{2}+(\bar\alpha-\frac{1}{2} )\ln x \right],\label{10a}
\end{eqnarray} where $N_C$ is a normalization constant. The derivative of
the wave function can be obtained from Eq. (\ref{10a}) and one obtains the
following log-derivative of the wave function in the asymptotic region
\begin{eqnarray} \lim_{x\to \infty} \frac{\phi'(x)}{\phi(x)}&=&
\left[-x+\left( \bar \alpha -\frac{1}{2}\right)\frac{1}{x} \right].
\label{11} \end{eqnarray} Equation (\ref{6}) is integrated numerically for
a given $\bar \alpha $ by the four-point Runge-Kutta rule starting at the
origin ($x=0$) with the initial boundary condition (\ref{10}) with a trial
$\phi'(0)$ in steps of $dx = 0.001$. Using Eq. (\ref{6}) the integration
is propagated to $x=x_{\mbox{max}}$, where the asymptotic condition
(\ref{11})  is valid. The agreement between the numerically calculated
log-derivative of the wave function and the theoretical result (\ref{11})
is enforced to four significant figures. The maximum value of $x$ up to
which we need to integrate to obtain this precision is $x_{\mbox{max}}=3$.
If for a trial $\phi'(0)$, this precision can not be obtained, a new value
of $\phi'(0)$ is to be chosen. The procedure is repeated until the
converged solution is obtained.

\section{Numerical Results} In our numerical study we would be interested
only in the case of attractive interaction (negative $\eta$). We consider
the experimentally relevant situation of $^7$Li \cite{3,4}.  In this case
the trap frequencies in the experiment of Ref. \cite{4} along the $X$,
$Y$, and $Z$ directions are 150.6 Hz, 152.6 Hz, and 131.5 Hz which lead to
a maximum of about 1400 trapped atoms.  The deviation from spherical
symmetry in this case is small and in order to have a qualitative
understanding of the condensate we consider the trap to be spherically
symmetric with $N_{\mbox{max}}= 1400$. We reconfirm from the numerical
solution of the GP equation that \cite{15} \begin{equation}\label{12}
\frac{|\eta| N_{\mbox{max}}}{4}=0.575.  \end{equation} Considering the
known result for $^7$Li, that $N_{\mbox{max}}=1400$ \cite{4}, we obtain
from Eq. (\ref{12}), $\eta = -0.00164$ and we use this value of $\eta$ in
our numerical calculation.  In our calculation we use three values of $N$,
e.g., $N= 1300, 1000,$ and 500 ($N < N_{\mbox{max}}=1400$). First we
calculate the critical temperatures in these cases from the
number equation (\ref{8}) and obtain the values $T_0 = 10.27
\hbar \omega/k_B, 9.41 \hbar \omega/k_B, 7.47 \hbar \omega/k_B$ for
  $N=1300, 1000$ and 500, respectively.

In the case of $^7$Li, the estimated value of $\eta (=-$0.00164) is quite
small. This corresponds  to a very weak coupling and we find that the
lowest-order solution is graphically almost indistinguishable from the
converged
solution for the condensate fraction, chemical potential,  and
total 
energy.
The estimated
coupling $\eta$ for $^{87}$Rb is 0.0248
\cite{1a,1b,11}, which is more than ten times larger in magnitude than the
coupling for $^7$Li. In the case of $^{87}$Rb, already the lowest order
result was very good \cite{11,15}. Hence the very rapid convergence of the
present results is not quite unexpected. In the present study we only
exhibit the converged result after two iterations.

\vskip -3.8cm
\postscript{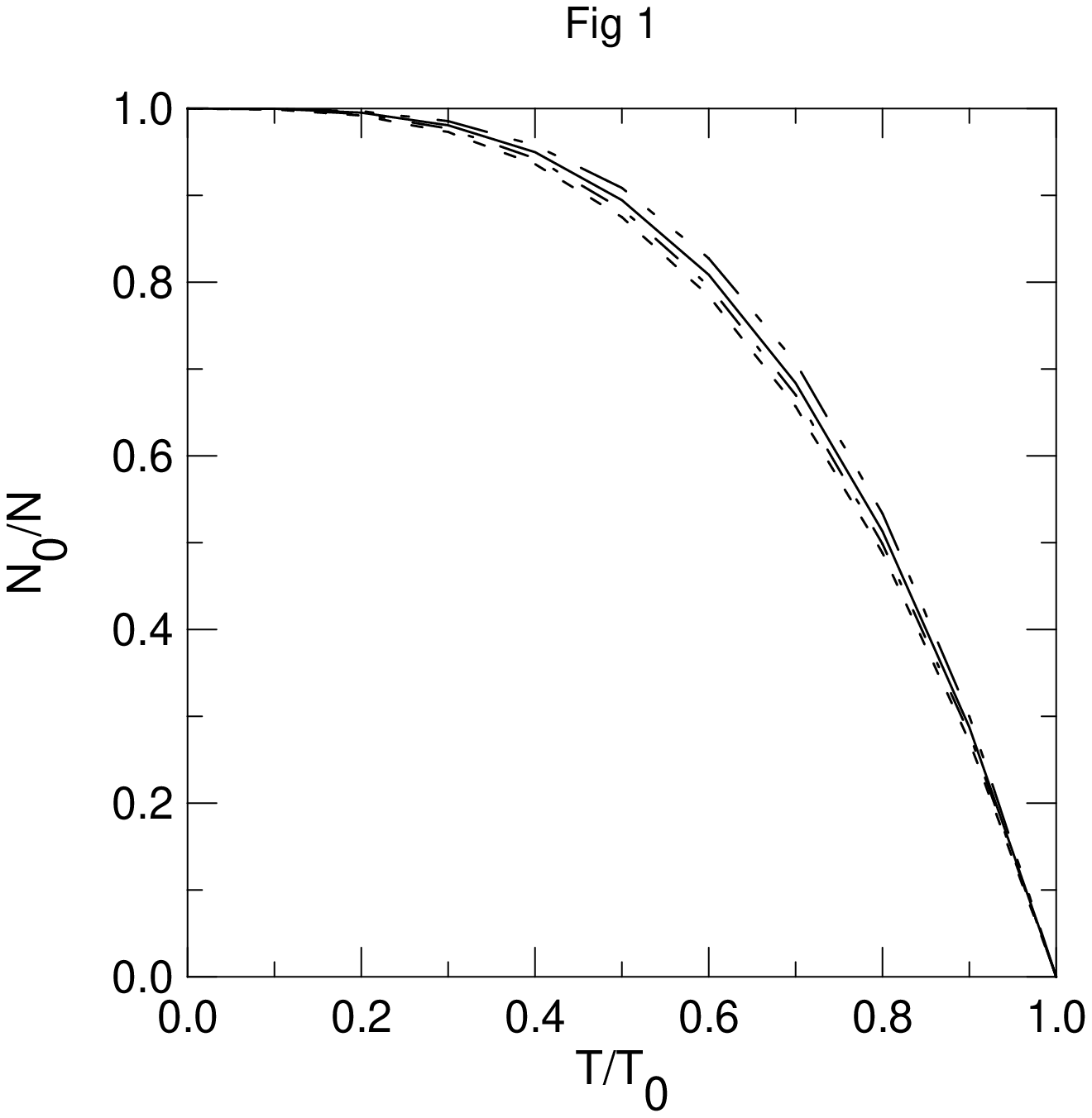}{1.0}    
\vskip -2.1cm

\vskip .4cm

{ {\bf Fig. 1.}  Converged
 condensate fraction $N_0/N$ as a function of $T/T_0$ for  $\eta =-
0.00164$
and  $N= $ 1300 (full line) and 500 (dashed-dotted line);
for $\eta=-0.01$ and $N=200$ (dashed-double-dotted line); and
for trapped ideal Bose gas (dotted line).  }

In Fig. 1 we present the temperature ($T/T_0$) dependence of the
condensate fraction
$N_0/N$ for $^7$Li with $\eta = - 0.00164$ for $N=1300$ and 500. The
result for $N=1000$ is indistinguishable from that of $N=1300$.
The result for $\eta =0$ corresponding to  an ideal Bose gas in
a harmonic trap is also shown in this figure. For comparison, the result
in the case of  a stronger attractive interaction for $\eta = -0.01$ and
$N=200$ is also shown.
In the
case of repulsive interaction,
the result for $N/N_0$ at a particular
temperature $T/T_0$ is smaller than that for an ideal Bose gas \cite{15}.
In the
present case of attractive interaction, the result for $N/N_0$ at a
particular
temperature $T/T_0$ is larger than that for the ideal Bose gas. We also
find from this figure that, as expected, the result for the stronger
attractive interaction ($\eta = -0.01$) deviates more from the ideal gas
result than in the case of   $^7$Li ($\eta = -0.00164$). We did not
consider a much stronger attractive coupling, as because of Eq. (\ref{12})
this would correspond to an unacceptably small value for the number of
particles in the condensate. This number is already small for the case
$\eta = -0.01$ considered in this work.

As the solution of the GP equation (\ref{6}) in this case is nontrivial,
we show in Fig. 2 the wave functions $\phi(x)/x$ for  $N=1000$ at
temperatures $ T/T_0$ = 0, 0.4, 0.6, 0.8, and 0.9. As temperature
decreases, the wave function is more pronounced corresponding to an
increase in the number of particles $N_0$ in the condensate given by the
normalization $N_0 = (4/|\eta|)\int \phi^2(x) dx.$ For other values of
$N$, the wave functions are similar to those in Fig. 2 and we do not
show these wave functions here.

\vskip -3.8cm
\postscript{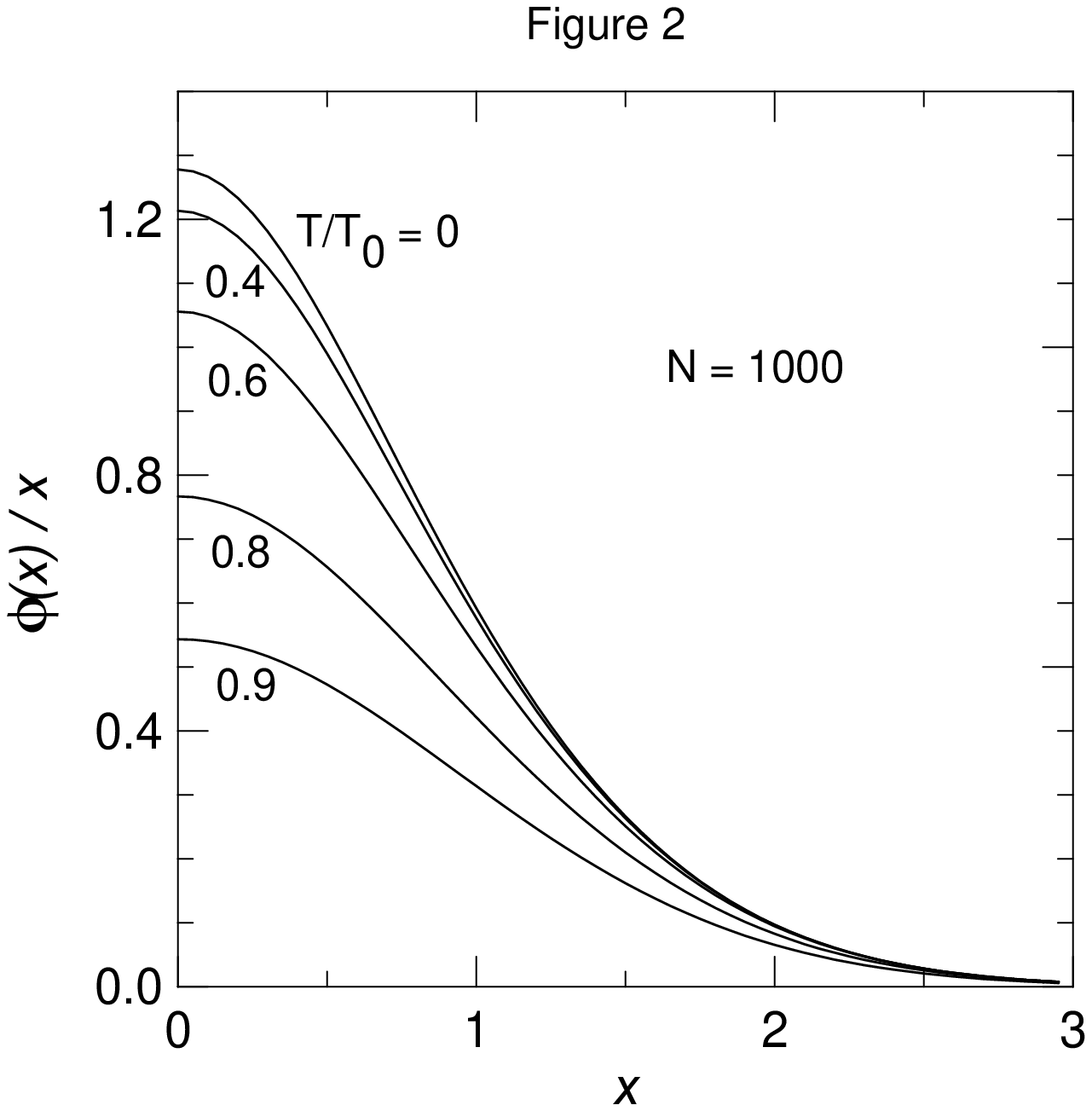}{1.0}    
\vskip -2.1cm

\vskip .4cm{\bf Fig. 2.}   GP wave function $\phi(x)/x$ of Eq. (\ref{6}) for $N=
1000$ at different
temperatures $T/T_0$ for $\eta = -0.00164$.

\vskip -3.8cm
\postscript{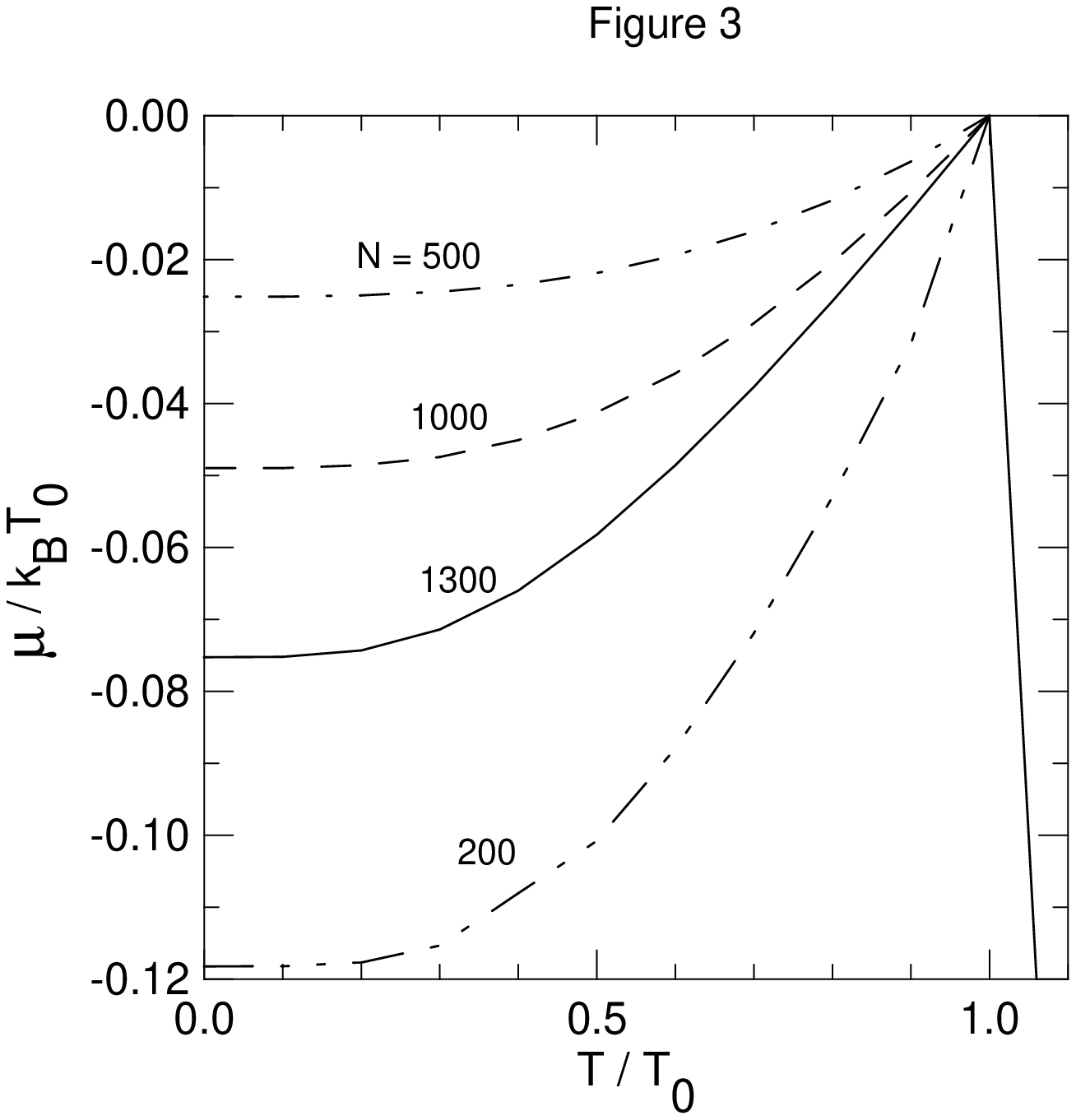}{1.0}    
\vskip -2.1cm

\vskip .4cm
{ {\bf Fig. 3.} 
Chemical potential
 $\mu/(k_BT_0)$ as a function of $T/T_0$ for  $\eta =-
0.00164$
and  $N= $ 1300 (full line), 1000 (dashed line) and 500 (dashed-dotted
line); and for $\eta = -0.01$ and $N=200$ (dashed-double-dotted line).}

In Fig. 3 we show the chemical potential of the system at different
temperatures $T/T_0$ for $N=1300, 1000,$ and 500 and $\eta = -0.00164$.
For comparison we also show the result for the stronger attractive
interaction with $N=200$ and $\eta =-0.01$. Above the critical temperature
$T>T_0$, the chemical potential for a trapped ideal Bose gas is negative
and it becomes zero at the critical temperature \cite{15}.  The same is
true in the present simplified model where the very weakly interacting
noncondensed gas is taken to be noninteracting. The only effect of
interaction is considered via the condensate. The effect of the
interaction among the atoms of the noncondensed gas could be important for
$T$ just below $T_0$ when there will be a large fraction of noncondensed
gas and a small fraction of condensed gas. For $T$ close to 0 when most
atoms are condensed such effect could be neglected.  At present a correct
description of BEC thermodynamics including the interaction among
noncondensed atoms is beyond the scope of the present work. Hence,
although it would be more appropriate to treat the noncondensed gas to be
interacting, the effect of interaction in the treatment of the
noncondensed gas is expected to be negligibly small in the present study
of very weakly interacting Bose gas ($\eta= -0.0164$)  and is neglected.
In the case of the ideal Bose gas, below the critical temperature the
chemical potential is identically equal to zero. For the present case of
the trapped Bose gas with attractive interaction, the chemical potential,
after becoming zero at $T=T_0$ from a negative value, becomes negative
again as the temperature is reduced below the critical temperature. The
chemical potential for the stronger attractive interaction ($\eta =-0.01$)
deviates more from the trapped ideal gas result ($\mu = 0$) below the
critical temperature, than in the case of the weak attractive interaction
of $^7$Li ($\eta = -0.00164$).  In the case of repulsive interaction, the
chemical potential becomes positive for temperatures below $T=T_0$
\cite{15}.

\vskip -3.8cm
\postscript{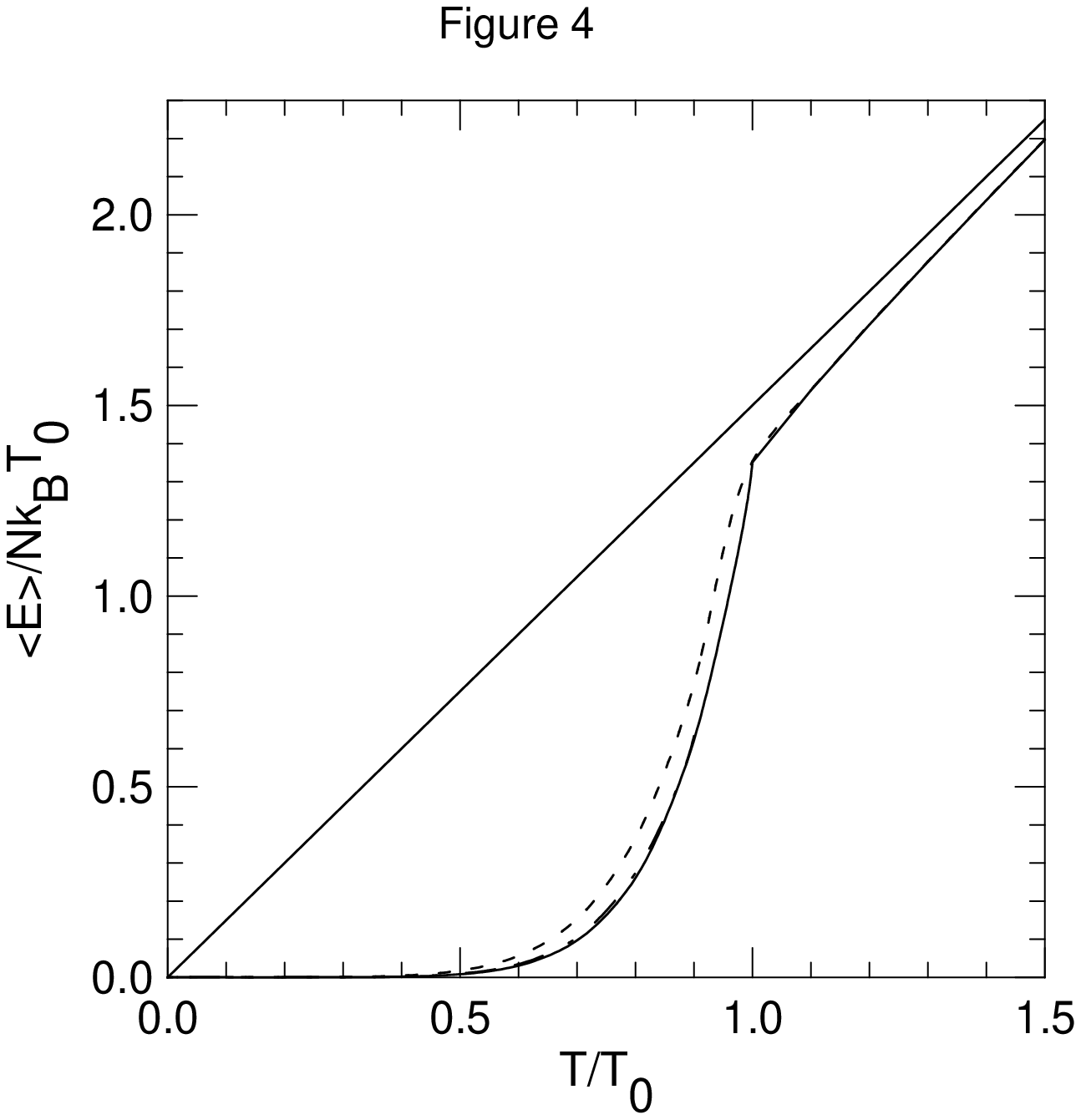}{1.0}    
\vskip -2.1cm

\vskip .4cm

{ {\bf Fig. 4.}    
 Energy  $\langle E \rangle/(Nk_BT_0)$ as a function of
$T/T_0$ for $\eta = -0.00164$ and
$N =1300$ (full line)  and 500 (dashed-dotted line);
for $\eta = 0.00164$ and
$N =1300$
(dotted line);
 and the classical Maxwell-Boltzmann
distribution (straight line). }
\vskip .2cm

In Fig. 4 we plot the temperature dependence of  energy $\langle E
\rangle/(Nk_BT)$ for $\eta =-0.00164$ and $N =1300$ and 500.  These two
cases lead to almost identical energies as can be seen from Fig. 4.  For
comparison, in this figure we also show the result for the repulsive
interaction for $\eta = 0.00164$ and $N =1300$. The energy for the trapped
ideal Bose gas should lie between the energies for the attractive and
repulsive cases mentioned above. The classical Maxwell-Boltzmann result is
also shown. For Bose-Einstein condensation to materialize the energy of
the system should be lower that the classical result. The energy in the
attractive case is smaller than the corresponding repulsive case below
critical temperature \cite{11,15}.

\section{Conclusion}

In conclusion, we studied the temperature dependence of condensate
fraction, chemical potential, and total energy for a trapped $^7$Li gas
consisting of 1300, 1000, and 500 atoms with attractive interaction using
a mean-field two-fluid model. The
maximum number of atoms allowed in this case is  1400 \cite{3,4}.
We employed an iterative solution scheme of the system
of equations of this model as in Refs. \cite{11,15}.
In the case of $^7$Li the attractive interaction is very weak
and the the system of equations
leads to rapid convergence.
The condensate was
described by the converged numerical solution of the Gross-Pitaevskii
equation \cite{9a,9b,10a,10b}.  As the interaction is weak, the results
for 
condensate
fraction, chemical potential, and  energy of $^7$Li are very close
to the corresponding results for the trapped ideal Bose gas. However, the
deviation from the result of the ideal Bose gas in case of $^7$Li is in
the opposite direction compared to the corresponding deviation in case of
repulsive interaction \cite{11,15}. For example, in all cases the chemical
potential is zero at the critical temperature. For ideal Bose gas it
continues to be zero below the critical temperature.
Below the critical
temperature the chemical potential turns negative  in case of $^7$Li,
whereas it turns positive in case of $^{87}$Rb \cite{15}  where
the interaction is repulsive. Although, in the present attractive case the
number of particles in the condensate is small, the present results for
the thermodynamic observables are quite reasonable physically. This
demonstates   that the mean-field two-fluid models \cite{11,13,14,15} used
to study the
thermodynamic observables for  the BE condensate in the repulsive case are
quite useful in the attractive case also.

We thank Dr. A. Gammal for useful discussion. The work is supported in
part by the Conselho Nacional de
Desenvolvimento
Cient\'\i fico e Tecnol\'ogico and Funda\c c\~ao de Amparo
\`a Pesquisa do Estado de S\~ao Paulo of Brazil.


\begin{references} 

\bibitem{1a}M. H. Anderson, J. R. Ensher, M. R. Mathews,
C. E. Wieman,  E. A. Cornell, Science { 269} (1995) 198.

\bibitem{1b}J. R.
Ensher, D. S. Jin, M. R. Mathews, C. E. Wieman,  E. A.  Cornell, {Phys.
Rev. Lett.} { 77} (1996) 4984.

\bibitem{2}K. B. Davis, M. O. Mewes, M. R. Andrews, N. J. van Druten, D.
S. Durfee, D. M. Kurn,  W. Ketterle, {Phys. Rev. Lett.} { 75}
(1995) 3969.

\bibitem{3}C. C. Bradley, C. A. Sackett, J. J. Tollett,  R. G. Hulet,
Phys. Rev. Lett.  { 75} (1995) 1687.

\bibitem{4}C. C. Bradley, C. A. Sackett,  R. G. Hulet,
Phys. Rev. Lett. { 78} (1997) 985.

\bibitem{5}D. G. Fried, T. C. Killian, L. Willmann, D. Landhuis, S. C.
Moss, D. Kleppner, T. J. Greytak, {Phys. Rev. Lett.} { 81} (1998) 3811.


\bibitem{6a}E. P. Gross, Nuovo Cimento { 20} (1961) 454.
\bibitem{6b}  L. P.
Pitaevskii,
Zh. Eksp. Teor. Fiz. { 40} (1961) 646 [Sov. Phys. JETP { 13} (1961) 451].

\bibitem{7a}C. A. Sackett, H. T. C. Stoof,  R. G. Hulet, {Phys. Rev.
Lett.} {80}  (1998)  2031.
\bibitem{7b}  M. Houbiers, H. T. C.  Stoof, Phys.
Rev. A 54 (1996)  5055.

\bibitem{7c}  H. T. C.  Stoof,
Phys. Rev. A  49 (1994) 3824.


\bibitem{8a} F. Dalfovo,  S. Stringari, Phys. Rev. A { 53} (1996) 2477.
\bibitem{8b}
R. J. Dodd, M. Edwards, C. J. Williams, C. W. Clark, M. J. Holland, P. A.
Ruprecht,  K. Burnett, Phys. Rev. A { 54} (1996)  661.

\bibitem{9a}M. Edwards,  K. Burnett, Phys. Rev. A { 51} (1995) 1382.
\bibitem{9b}  P.
A. Ruprecht, M. J. Holland, K. Burnett,  M. Edwards, Phys. Rev. A { 51}
(1995) 4704.

\bibitem{10a} A. Gammal, T. Frederico,  L. Tomio, Phys. Rev. E {60}
(1999)  2421.
\bibitem{10b} S. K. Adhikari, Phys. Lett. A 265 (2000) 91.

\bibitem{11} A. Minguzzi, S. Conti,  M. P. Tosi, J. Phys.: Condens. Matter
{ 9} (1997) L33.

\bibitem{13} S. Giorgini, L. P. Pitaevskii,  S. Stringari, Phys. Rev. A {
 54} (1996) R4633.

\bibitem{13a1}S. Giorgini, L. P. Pitaevskii,  S. Stringari, J. Low Temp.
Phys. { 109} (1997) 309.
\bibitem{13a2}  H. Shi,  W. M. Zheng, Phys. Rev. A { 56}
(1997) 1046.
 
\bibitem{14}V. N. Popov, ``Functional Integrals and Collective
Excitations", (Cambridge University Press, Cambridge, 1987).


\bibitem{15} F. Dalfovo, S. Giorgini, L. P. Pitaevskii,  S. Stringari,
Rev. Mod.  Phys. {71} (1999) 463.


\bibitem{16}L. V. Hau, B. D. Busch, C. Liu, Z. Dutton, M. M. Burns,
J. A. Golovchenko, Phys. Rev. A { 58} (1998) R54.

\bibitem{12a}M. Bayindir, B. Tanatar,  Z. Gedik, Phys. Rev. A { 59}
(1999) 1468.
\bibitem{12b}  M. Bayindir,  B. Tanatar, Phys. Rev. A { 58} (1998) 3134.



\bibitem{17} V. Bagnato, D. E. Pritchard, D. Kleppner,
Phys. Rev. A { 35} (1987) 4354. 









\end{references}
\end{document}